\documentclass{llncs}

\usepackage{makeidx}  
\usepackage{epsfig}
\usepackage{stmaryrd}
\usepackage{hyperref}
\usepackage{amsmath}
\usepackage{latexsym}
\usepackage{ amssymb }
\usepackage{paralist}
\usepackage{times}

\renewcommand{\vec}[1]{\overline{#1}}

\newcommand{\names}{\mathrm{names}}
\newcommand{\kw}[1]{{\texttt{\small #1}}}
\newcommand{\boundp}{\kw{BOUND}}
\newcommand{\andq}{\mathrel{\kw{.}}}
\newcommand{\optq}{\mathrel{\kw{OPT}}}
\newcommand{\unionq}{\mathrel{\kw{UNION}}}
\newcommand{\filterq}{\mathrel{\kw{FILTER}}}
\newcommand{\graphq}[2]{\kw{GRAPH}~#1~\{#2\}}
\newcommand{\selectq}[2]{\kw{SELECT}~#1~\kw{WHERE}~#2}
\newcommand{\constructq}[2]{\kw{CONSTRUCT}~#1~\kw{WHERE}~#2}

\newcommand{\QSrc}[2]{\mathcal{S}\SB{#1}_{#2}}
\newcommand{\PSrc}[2]{\mathcal{S}\SB{#1}_{#2}}
\newcommand{\insertdata}[1]{\kw{INSERT}~\kw{DATA}~\{#1\}}
\newcommand{\deletedata}[1]{\kw{DELETE}~\kw{DATA}~\{#1\}}
\newcommand{\insertinto}[2]{\kw{INSERT}~\{#1\}~\kw{WHERE}~#2}
\newcommand{\deletefrom}[2]{\kw{DELETE}~\{#1\}~\kw{WHERE}~#2}
\newcommand{\deleteinsert}[3]{\kw{DELETE}~\{#1\}~\kw{INSERT}~\{#2\}~\kw{WHERE}~#3}
\newcommand{\deletewhere}[1]{\kw{DELETE}~\kw{WHERE}~\{#1\}}
\newcommand{\loadinto}[2]{\kw{LOAD}~#1~\kw{INTO}~#2}
\newcommand{\clear}[1]{\kw{CLEAR}~#1}
\newcommand{\creategraph}[1]{\kw{CREATE}~#1}
\newcommand{\dropgraph}[1]{\kw{DROP}~#1}
\newcommand{\addto}[2]{\kw{ADD}~#1~\kw{TO}~#2}
\newcommand{\moveto}[2]{\kw{MOVE}~#1~\kw{TO}~#2}
\newcommand{\copyto}[2]{\kw{COPY}~#1~\kw{TO}~#2}

\newcommand{\Eval}[2]{\SB{#1}_#2}
\newcommand{\compat}{\mathrel{\mathsf{compat}}} 
\newcommand{\QEval}[2]{\SB{#1}_{#2}}
\newcommand{\PEval}[2]{\SB{#1}_{#2}}

\newcommand{\SB}[1]{\llbracket #1 \rrbracket}

\newcommand{\rdf}[3]{\langle #1~#2~#3\rangle}
\newcommand{\lit}{\ell}
\newcommand{\iri}{\iota}
\newcommand{\var}[1]{{?}#1}

\newcommand{\pfn}{\rightharpoonup}
\newcommand{\D}{\mathcal{D}}
\newcommand{\Lit}{\mathsf{Lit}}
\newcommand{\IRI}{\mathsf{Id}}
\newcommand{\Atom}{\mathsf{Atom}}
\newcommand{\Var}{\mathsf{Var}}

\newcommand{\dom}{\mathrm{dom}}

\newcommand{\true}{\kw{true}}
\newcommand{\false}{\kw{false}}
\newcommand{\error}{\kw{error}}

\begin{document}
%
%
\title{Dynamic Provenance for SPARQL Updates}
\titlerunning{Dynamic Provenance for SPARQL Updates}  
%
%
\author{Harry Halpin \and James Cheney}

\authorrunning{Halpin and Cheney} 
 \institute{World Wide Web Consortium/MIT, 32 Vassar Street
 Cambridge, MA 02139 USA \\
 \email{hhalpin@w3.org},
 \and
 University of Edinburgh,
 School of Informatics,\\
 10 Crichton St.,\\
 Edinburgh UK EH8 9AB}

\maketitle              

\begin{abstract}
While the Semantic Web currently can exhibit provenance information by using the W3C PROV standards, there is a ``missing link'' in connecting PROV to storing and querying for dynamic changes to RDF graphs using SPARQL.  Solving this problem  would be required for such clear use-cases as the creation of version control systems for RDF. While some provenance models and annotation techniques for storing and querying provenance data originally developed with databases or workflows in mind transfer readily to RDF and SPARQL, these techniques do not readily adapt to describing changes in dynamic RDF datasets over time.   
In this paper we explore how to adapt the dynamic copy-paste
provenance model of Buneman et al.~\cite{buneman06sigmod} to RDF
datasets that change over time in response to SPARQL updates, how to
represent the resulting provenance records themselves as RDF in a manner compatible with W3C PROV, and how the
provenance information can be defined by reinterpreting SPARQL
updates.
The primary contribution of this paper is a semantic framework that enables the semantics of SPARQL Update to be used as the basis for a `cut-and-paste' provenance model in a principled manner. 
\end{abstract}

\section{Introduction}

It is becoming increasingly common to publish scientific and governmental data on the Web as RDF (the Resource Description Framework, a W3C standard for structured data on the Web) and to attach provenance data to this information using the W3C PROV standard.  In doing so, it is crucial to track not only the provenance metadata, but the changes to the graph itself, including both its derivation process and a history of changes to the data over time.  Being able to track changes to RDF graphs could, in combination with the W3C PROV standards and SPARQL, provide the foundation for addressing the important use-case of creating a Git-like version control system for RDF.  

The term provenance is used in several different ways, often aligning with research in two different communities, in particular the database community and the scientific workflow community. We introduce some  terminology to distinguish different uses of the term.  There is a difference between \emph{static} provenance that describes data at a given point in time versus \emph{dynamic} provenance that describes how artifacts have evolved over time. Second, there is a difference between provenance for \emph{atomic} artifacts that expose no internal structure as part of the provenance record, versus provenance for \emph{collections} or other structured artifacts. The workflow community has largely focused on static provenance for atomic artifacts, whereas much of the work on provenance in databases has focused on dynamic provenance for collections (e.g. tuples in relations). An example of static provenance would be attaching the name of an origin and time-stamp to a set of astronomical RDF data. Thus static provenance can often be considered  metadata related to provenance. Currently the W3C PROV data model and vocabulary \cite{prov-ontology}, provides a standard way to attach this provenance, and other work such  as PROV-AQ \cite{prov-access} provides options for extracting this metadata by virtue of HTTP.

 An example of \emph{dynamic provenance} would be given by a single astronomical data-set that is updated over time, and then if some erroneous data was added at a particular time, the data-set could be queried for its state at a prior time before the erroneous data was added so the error could be corrected. Thus dynamic provenance can in some cases be reduced to issues of history and version control. Note that static and dynamic  provenance are not orthogonal, as we can capture \emph{static} provenance metadata at every step of recording dynamic provenance. The W3C Provenance Working Group has focused primarily on static provenance, but we believe a mutually beneficial relationship between the W3C PROV static provenance and SPARQL Update with an improved provenance-aware semantic model will allow dynamic provenance capabilities to be added to RDF. 
 While fine-grained dynamic provenance imposes overhead that may make
 it unsuited to some applications, there are use-cases where knowing
 exactly when a graph was modified is necessary for reasons of
 accountability, including data-sets such as private financial and
 public scientific data. 

\subsection{Related Literature}

The workflow community has largely focused on declaratively describing causality or derivation steps of processes to aid repeatability for scientific experiments, and these requirements have been a key motivation for the Open Provenance Model (OPM)~\cite{opm11,moreau11jws}, a vocabulary and data model for describing processes including (but certainly not limited to) runs of scientific workflows.  OPM is important in its own right, and has become the foundation for the W3C PROV data model \cite{prov-dm}. The formal semantics of PROV have been formalized, but do not address the relationship between the provenance and the semantics of the processes being described~\cite{prov-sem}. 
However, previous work on the semantics of OPM, PROV, and other
provenance vocabularies focuses on temporal and validity
constraints~\cite{eps271819} and does not address the meaning
of the processes being represented --- one could in principle use them
either to represent static provenance for processes that construct new
data from existing artifacts or to represent dynamic provenance that
represents how data change over time at each step, such as needed by version control systems.
Most applications of OPM seem to have focused on static provenance,
although PROV explicitly grapples with issues concerning representing
the provenance of objects that may be changing over time, but does not
provide a semantics for storing and querying changes over time. Our
approach to dynamic provenance is complementary to work on OPM and
PROV, as we show how any provenance metadata vocabulary such as the
W3C PROV ontology \cite{prov-ontology} can be connected directly to
query and update languages by presenting a semantics for a generalized
provenance model and showing how this builds upon the formal semantics
of the query and update languages. 

Within database research, complex provenance is also becoming increasingly seen as necessary, although the work has taken a different turn than that of the workflow community. Foundational work on provenance for database queries distinguishes between \emph{where-provenance}, which is the ``locations in the source databases from which the data was extracted,'' and \emph{why-provenance}, which is ``the source data that had some influence on the existence of the data''~\cite{buneman01icdt}. Further, increasing importance is being placed on \emph{how-provenance}, the operations used to produce the derived data and other annotations, such as who precisely produced the derived data and for what reasons.  There is less work considering provenance for updates; previous work on provenance in databases has focused on simple atomic update operations: insertion, deletion, and
copy~\cite{buneman06sigmod}. A provenance-aware database should support queries that permit users to find the ultimate or proximate `sources' of some data and apply data provenance techniques to understand why a given part of the data was inserted or deleted by a given update. There is considerable work on correct formalisms to enable provenance in databases \cite{Green-2007}. This work on dynamic provenance is related to earlier work on version control in unstructured data, a classic and well-studied problem for information systems ranging from source code management systems to temporal databases and data archiving \cite{snodgrass}.  Version control systems such as CVS, Subversion, and Git have a solid track record of tracking versions and (coarse-grained) provenance for text (e.g. source code) over time and are well-understood in the form of temporal annotations and a log of changes. 

On the Semantic Web, work on provenance has been quite diverse.  There is widely implemented support for \emph{named graphs}, where each graph $G$ is identified with a \emph{name  URI}~\cite{carroll05jws}, and as the new RDF Working Group has standardized the graph name in the semantics, the new standards have left the practical use of such a graph name under-defined; thus, the graph name could be used to store or denote provenance-related information such as time-stamps. Tackling directly the version control aspect of provenance are proposals such as Temporal RDF~\cite{Gutierrez05temporalrdf} for attaching temporal annotations  and a generalized Annotated RDF for any partially-ordered sets already exist~\cite{udrea10tocl,lopes10iswc}. However, by confining provenance annotations to partially-ordered sets, these approaches do not allow for a (queryable) graph structure for more complex types of provenance that include work such as the PROV model.  Static provenance techniques have been investigated for RDFS inferences~\cite{buneman10swpm,flouris09iswc,vicky:iswc} over RDF datasets.  Some of this work considers updates, particularly Flouris et al.~\cite{flouris09iswc}, who consider the problem of how to maintain provenance information for RDFS inferences when tuples are inserted or deleted using coherence semantics. Their solution uses `colouring' (where the color is the URI of the source of each triple) and tracking implicit triples~\cite{flouris09iswc}.  



Understanding provenance for a language requires understanding the ordinary semantics of the language. Arenas et al. formalized the semantics of SPARQL~\cite{perez09tods,arenas10swim}, and the SPARQL Update recommendation proposes a formal model for updates~\cite{sparql-update}.  Horne et al.~\cite{horne11jist} propose an operational semantics for SPARQL Updates, which however differs in some respects from the SPARQL 1.1 standard and does not deal with named graphs.

\subsection{Overview}

In this paper, we build on Arenas et al.'s semantics of
SPARQL~\cite{perez09tods,arenas10swim} and extend it to formalize
SPARQL Update semantics (following a denotational approach similar to
the SPARQL Update Formal Model~\cite{sparql-update}). Then, as our
main contribution, we detail a provenance model for SPARQL queries and
updates that provides a (queryable) record of how the raw data in a
dataset has changed over time.
This change history includes  a way to insert static provenance metadata using a full-scale provenance ontology such as PROV-O~\cite{prov-ontology}.

Our hypothesis is that a simple vocabulary, composed of insert, delete, and copy operations as introduced by Buneman et al.~\cite{buneman06sigmod}, along with explicit identifiers for update steps, versioning relationships, and metadata about updates provides a flexible format for dynamic provenance on the Semantic Web. A primary advantage of our methodology is it keeps the changes to raw data separate from the changes in provenance metadata, so legacy applications will continue to work and the cost of storing and providing access to provenance can be isolated from that of the raw data.  We will introduce the semantics of SPARQL queries, then  our semantics for SPARQL updates, and finally describe our semantics for dynamic provenance-tracking for RDF graphs and SPARQL updates. To summarize, our contributions are an extension to the  semantics of SPARQL Update that includes provenance semantics to handle dynamic semantics~\cite{perez09tods,arenas10swim} and a vocabulary for representing changes to RDF graphs made by SPARQL updates, and a translation from ordinary SPARQL updates to provenance-aware updates that record provenance as they execute.

\section{Background: Semantics of SPARQL Queries}

We first review a simplified (due to space limits) version of Arenas
et. al.'s formal semantics of SPARQL \cite{perez09tods,arenas10swim}.
Note that this is not original work, but simply a necessary precursor
to the semantics of SPARQL Update and our extension that adds
provenance to SPARQL Update. The main simplification is that we
disallow nesting other operations such as $\andq,\unionq$, etc.~inside
$\graphq{A}{\ldots}$ patterns. This limitation is inessential.

Let $\Lit$ be a set of literals (e.g. strings), let $\IRI$ be a set of resource identifiers, and let $\Var$ be a set of variables usually written $\var{X}$.  We write $\Atom = \Lit \cup \IRI$ for the set of atomic values, that is literals or ids.  The syntax of a core algebra for SPARQL discussed in~\cite{arenas10swim} is as follows:
\begin{eqnarray*}
  A &::=& \lit \in \Lit \mid \iri  \in \IRI \mid \var{X} \in \Var\\
  t &::=& \rdf{A_1}{A_2}{A_3}\\
  C &::=& \{t_1 ,\ldots,t_n\} \mid \graphq{A}{t_1,\ldots,t_n}  \mid C ~C'\\
  R &::=& \boundp(?x) \mid A = B \mid R \wedge R' \mid R \vee R'  \mid \neg R\\
  P &::=& C \mid P \andq P' \mid P \unionq P' \mid P \optq P' \mid P \filterq R \\
  Q &::=& \selectq{\var{\vec{X}}}{P}  \mid \constructq{C}{P}
\end{eqnarray*}

Here, $C$ denotes basic graph (or dataset) patterns that may contain variables; $R$ denotes conditions; $P$ denotes patterns, and $Q$ denotes queries.  We do not distinguish between subject, predicate and object components of triples, so this is a mild generalization of~\cite{arenas10swim}, since SPARQL does not permit literals to appear in the subject or predicate position or as the name of a graph in the $\graphq{A}{P}$ pattern, although the formal semantics of RDF allows this and the syntax may be updated in forthcoming work on RDF.  We also do not consider blank nodes, which pose complications especially when updates are concerned, and we instead consider them to be skolemized (or just replaced by generic identifiers), as this is how most implementations handle blank nodes.  There has been previous work giving a detailed treatment of the problematic nature of blank nodes and why skolemization is necessary in real-world work. 

The semantics of queries $Q$ or patterns $P$ is defined using
functions from \emph{graph stores} $\D$ to sets of \emph{valuations}
$\mu$. A graph store $\D = (G,\{g_i\mapsto G_1\ldots,g_n\mapsto
G_n\})$ consists of a default graph $G_0$ together with a mapping from
names $g_i$ to graphs $G_i$. Each such graph is essentially just a set
of ground triples. We often refer to graph stores as \emph{datasets},
although this is a slight abuse of terminology.

We overload set operations for datasets, e.g. $\D \cup \D'$ or $\D
\setminus \D'$ denotes the dataset obtained by unioning or
respectively subtracting the default graphs and named graphs of $\D$
and $\D'$ pointwise. If a graph $g$ is defined in $\D$ and undefined
in $\D'$, then $(\D \cup \D')(g) = \D(g)$ and similarly if $g$ is
undefined in $\D$ and defined in $\D'$ then $(\D \cup \D')(g) =
\D'(g)$; if $g$ is undefined in both datasets then it is undefined in
their union. For set difference, if $g$ is defined in $\D$ and
undefined in $\D'$ then $(\D \setminus \D')(g) =\D(g)$; if $g$ is
undefined in $\D$ then it is undefined in $(\D \setminus \D')$.
Likewise, we define $\D \subseteq \D'$ as $\D' = \D \cup \D'$.

A \emph{valuation} is a partial function $\mu : \Var \pfn \Lit \cup
\IRI$.  We lift valuations to functions $\mu : \Atom \cup \Var \to
\Atom$ as follows:
\begin{eqnarray*}
   \mu(\var{X} ) &=&\mu(X)\\
   \mu(a) &=& a \qquad a \in \Atom
\end{eqnarray*}
that is, if $A$ is a variable $\var{X}$ then $\mu(A) = \mu(X)$ and
otherwise if $A$ is an atom then $\mu(A) = A$.  We thus consider all
atoms to be implicitly part of the domain of $\mu$.  Furthermore, we
define $\mu$ applied to triple, graph or dataset patterns as follows:
\begin{eqnarray*}
  \mu(\rdf{A_1}{A_2}{A_3}) &=& \rdf{\mu(A_1)}{\mu(A_2)}{\mu(A_3)}\\
  \mu(\{t_1,\ldots,t_n\}) &=& (\{\mu(t_1),\ldots,\mu(t_n)\},\emptyset)\\
  \mu(\graphq{A}{t_1,\ldots,t_n}) &=& (\emptyset, \{\mu(A) \mapsto
  \{\mu(t_1),\ldots,\mu(t_n)\}\})\\
  \mu(C~C') &=& \mu(C) \cup \mu(C')
\end{eqnarray*}
where, as elsewhere, we define $\D \cup \D'$ as pointwise union of datasets.

The conditions $R$ are interpreted as three-valued formulas over the
lattice $L = \{\true,\false,\error\}$, where $\false
\leq \error \leq \true$, and $\wedge$ and $\vee$ are minimum and
maximum operations respectively, and $\neg \true = \false$, $\neg
\false = \true$, and $\neg \error = \error$.  The semantics of a condition is
defined as follows:
\begin{eqnarray*}
  \SB{\boundp(\var{X})}\mu &= & \left\{
    \begin{array}{ll}
      \false & \text{if }\var{X} \not\in \dom(\mu) \\
      \true & \text{if }\var{X} \in \dom(\mu) 
    \end{array}
  \right.\\
  \SB{A = B}\mu &=&  \left\{\begin{array}{ll}
      \error & \text{if }\text{$\{A,B\} \not\subseteq \dom(\mu)$}\\
      \true & \text{if }\mu(A) = \mu(B) \text{ where } A,B \in \dom(\mu)\\
      \false &  \text{if }\mu(A) \neq \mu(B) \text{ where } A,B \in \dom(\mu)
    \end{array}
  \right.\\
  \SB{\neg R}\mu &=& \neg \SB{R}\mu\\
  \SB{R \wedge R'}\mu &=& \SB{R}\mu \wedge \SB{R'}\mu\\
  \SB{R \vee R'}\mu &=& \SB{R}\mu \vee \SB{R'}\mu
\end{eqnarray*}
We write $\mu \models R$ to indicate that $\SB{R} = \true$.

We say that two valuations $\mu,\mu'$ are \emph{compatible} (or write $\mu \compat \mu'$) if for all variables $x \in \dom(\mu) \cap \dom(\mu')$, we have
$\mu(x) = \mu'(x)$.  Then there is a unique valuation $\mu \cup \mu'$ that behaves like $\mu$ on $\dom(\mu)$ and like $\mu'$ on $\dom(\mu')$.  We define the following operations on sets of valuations $\Omega$.
\begin{eqnarray*}
  \Omega_1 \Join \Omega_2 &=& \{\mu_1 \cup \mu_2 \mid \mu_1 \in
  \Omega_1, \mu_2 \in \Omega_2, \mu_1 \compat \mu_2 \}\\
\Omega_1 \cup \Omega_2 &=& \{ \mu \mid \text{$\mu \in \Omega_1$ or $\mu
  \in \Omega_2$}\}\\
\Omega_1 \setminus \Omega_2 &=& \{\mu \in \Omega_1\mid \not\exists
\mu' \in \Omega_2. \mu \compat \mu'\}\\
\Omega_1 \rtimes \Omega_2 &=& (\Omega_1 \Join \Omega_2) \cup (\Omega_1
\setminus \Omega_2)
\end{eqnarray*}
Note that union is the same as ordinary set union, but difference is
not, since $\Omega_1 \setminus \Omega_2$ only includes valuations that
are incompatible with all those in in $\Omega_2$.

Now we can define the meaning of a pattern $P$ in a dataset $\D$ as a set of valuations $\PEval{P}{\D}$, as follows:
\begin{eqnarray*}
  \PEval{C}{\D} &=& \{\mu \mid \dom(\mu) = vars(C) \text{ and }
  \mu(C) \subseteq \D\} \\
  \PEval{P_1 \andq P_2}{\D} &=& \PEval{P_1}{\D} \Join
  \PEval{P_2}{\D}\\
  \PEval{P_1 \unionq P_2}{\D} &=& \PEval{P_1}{\D} \cup
  \PEval{P_2}{\D}\\
  \PEval{P_1 \optq P_2}{\D} &=& \PEval{P_1}{\D} \rtimes
  \PEval{P_2}{\D}\\
  \PEval{P \filterq R}{\D} &=& \{\mu \in \PEval{P}{\D} \mid \mu
  \models R\}
\end{eqnarray*}
Note that, in contrast to Arenas et al.'s semantics for SPARQL with
named graphs~\cite{arenas10swim}, we do not handle $\graphq{A}{\ldots}$
patterns that contain other pattern operations such as $\andq$ or
$\unionq$, and we do not keep track of the ``current graph''
$G$. Instead, since graph patterns can only occur in basic patterns,
we can build the proper behavior of pattern matching into the
definition of $\mu(C)$, and we select all matches $\mu$ such that
$\mu(C) \subseteq \D$ in the case for $\PEval{C}{\D}$.

Finally, we consider the semantics of selection and construction
queries.  A selection query has the form $\selectq{\var{\vec{X}}}{P}$
where $\var{\vec{X}}$ is a list of distinct variables.  It simply
returns the valuations obtained by $P$ and discards the bindings of
variables not in $\vec{X}$.  A construction query builds a new graph
or dataset from these results.  Note that in SPARQL such queries only
construct anonymous graphs; here we generalize in order to use
construction queries to build datasets that can be inserted or
deleted.
\begin{eqnarray*}
  \QEval{\selectq{\var{\vec{X}}}{P}}{\D} &=& \{\mu|_{\vec{X}} \mid \mu
  \in \PEval{P}{\D}\}\\
\QEval{\constructq{C}{P}}{\D} &=& \bigcup\{\mu(C) \mid \mu \in  \PEval{P}{\D}\}
\end{eqnarray*}
Here, note that $\mu|_{\vec{X}}$ stands for $\mu$ restricted to the
variables in the list $\vec{X}$.

  We omit discussion of the \texttt{FROM} components of queries (which  are used to initialize the graph store by pulling data in from  external sources) or of the other query forms \texttt{ASK}, and \texttt{DESCRIBE}, as they are described elsewhere~\cite{arenas10swim} in a manner coherent with our approach. 

\section{The Semantics of SPARQL Update}

We will describe the semantics of the core language for atomic updates, based
upon~\cite{sparql-update}:
\begin{eqnarray*}
  U &::=& \insertinto{C}{P} \mid \deletefrom{C}{P}\\
&\mid& \deleteinsert{C}{C'}{P} \mid \loadinto{g}{g'}
  \mid \clear{g} \\
&\mid&  \creategraph{g} \mid \dropgraph{g} \mid \copyto{g}{g'} \mid \moveto{g}{g'} \mid \addto{g}{g'}
\end{eqnarray*}
We omit the $\kw{INSERT}~\kw{DATA}$ and $\kw{DELETE}~\kw{DATA}$ forms since they are definable in terms of $\kw{INSERT}$ and $\kw{DELETE}$.

SPARQL Update  \cite{sparql-update} specifies that transactions consisting of multiple updates should be applied atomically, but leaves some semantic
questions unresolved, such as whether aborted transactions have to
roll-back partial changes.  It also does not specify whether updates
in a transaction are applied sequentially (as in most imperative
languages), or using a snapshot semantics (as in most database update
languages).  Both alternatives pose complications, so in this paper we
focus on transactions consisting of single atomic updates. 

We model a collection of named graphs as a dataset
$\D$, as for SPARQL queries.  We consider only a single graph in isolation here, and not the case of multiple named graphs that may be being updated concurrently.  The semantics of an update operation $u$ on
dataset $\D$ is defined as $\Eval{U}{\D}$.

The semantics of a SPARQL Update $U$ in dataset $\D$ is defined as follows:
\begin{eqnarray*}
 \Eval{\deletefrom{C}{P}}{\D} &=& \D \setminus \QEval{\constructq{C}{P}}{\D}\\
  \Eval{\insertinto{C}{P}}{\D} &=& \D \cup
  \QEval{\constructq{C}{P}}{\D} \\
\Eval{\deleteinsert{C}{C'}{P}}{\D} &=& (\D\setminus
\QEval{\constructq{C}{P}}{\D}) 
\\ && {} \cup \QEval{\constructq{C'}{P}}{\D}\\
  \Eval{\loadinto{g_1}{g_2}}{\D} &=& \D[g_2 := \D(g_1) \cup \D(g_2)]\\
  \Eval{\clear{g}}{\D} &=& \D[g:=\emptyset]\\
  \Eval{\creategraph{g}}{\D} &=& \D \uplus \{g \mapsto \emptyset\}\\
  \Eval{\dropgraph{g}}{\D} &=& \D[g := \bot]\\
  \Eval{\copyto{g_1}{g_2}}{\D} &=& \left\{\begin{array}{ll}
\D[g_2 := \D(g_1)] & \text{if $g_1 \neq g_2$}\\
\D & \text{otherwise}
\end{array}\right.\\
  \Eval{\moveto{g_1}{g_2}}{\D} &=& \left\{\begin{array}{ll}
\D[g_2 := \D(g_1), g_1 := \bot] & \text{if $g_1 \neq g_2$}\\
\D &\text{otherwise}
\end{array}\right.\\
  \Eval{\addto{g_1}{g_2}}{\D} &=& \D[g_2 := \D(g_1) \cup \D(g_2)]\\
\end{eqnarray*}
Here, $\D[g:=G]$ denotes $\D$ updated by setting the graph named $g$
to $G$, and $\D[g:=\bot]$ denotes $\D$ updated by making $g$
undefined, and finally $\D \uplus [g := G]$ denotes $\D$ updated by
adding a graph named $g$ with value $G$, where $g$ must not already be
in the domain of $\D$. Set-theoretic notation is used for graphs, e.g.
$G \cup G'$ is used for set union and $G \setminus G'$ for set
difference, and $\emptyset$ stands for the empty graph. Note that the
$\copyto{g}{g'}$ and $\moveto{g}{g'}$ operations have no effect if $g
= g'$.  Also, observe that we do not model external URI
dereferences, and the $\loadinto{g}{g'}$ operation (which allows $g$
to be an external URI) behaves exactly as $\addto{g}{g'}$ operation
(which expects $g$ to be a local graph name).


\section{Provenance semantics}

A single SPARQL update can read from and write to several named graphs
(and possibly also the default graph). For simplicity, we restrict
attention to the problem of tracking the provenance of updates to a
single (possibly named) RDF graph.  All operations may still use the
default graph or other named graphs in the dataset as sources.  The
general case can be handled using the same ideas as for a single
anonymous graph, only with more bureaucracy to account for versioning
of all of the named graphs managed in a given dataset.

A graph that records all the updates of triples from a given graph $g$ is considered a \emph{provenance graph} for $g$. For each operation, a \emph{provenance record} is stored that track of the state of the graph at any given moment and their associated metadata. The general concept is that in a fully automated process  one should be able to re-construct the state of the given graph at any time from its provenance graph by following the SPARQL queries and metadata given in the provenance records for each update operation tracked. 

\begin{figure*}[tb]
\includegraphics[scale=0.4]{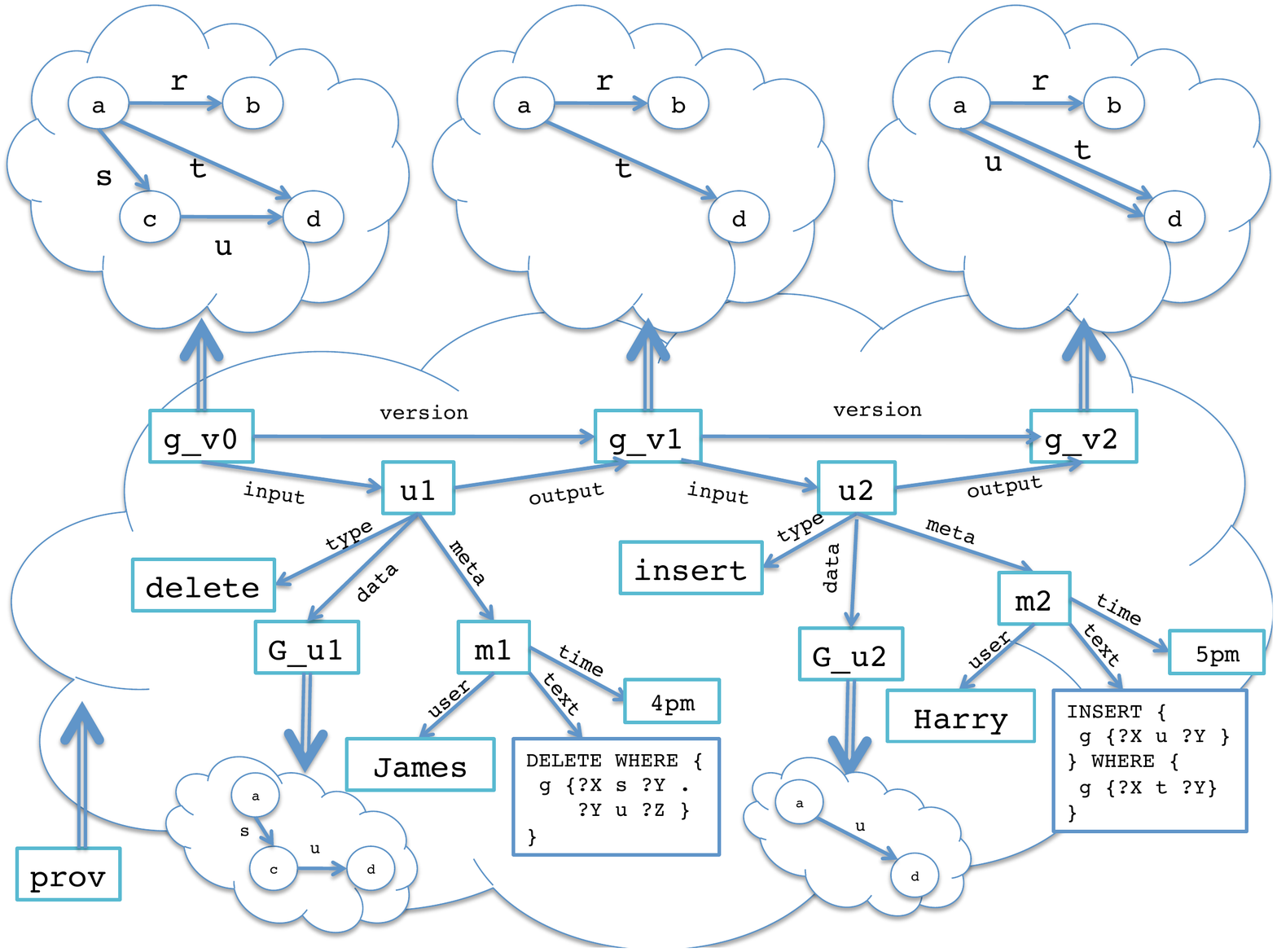}
\caption{Example provenance graph}\label{fig:example}
\end{figure*}

We model the provenance of a single RDF graph $G$ that is updated over
time as a set of history records, including the special provenance graph named
$prov$ which consists of indexed graphs for each operation such as \verb|G_v0|,\ldots,\verb|G_vn| and \verb|G_u1|\ldots,\verb|G_um|. These provenance records  are immutable; that is, once they have been created and initialized, the implementation should fixed so that their content cannot be changed. The index of all provenance records then is also strictly linear and consistent (i.e. non-circular), although branching could be allowed.  They can be stored as a special immutable type in the triple-store.  Intuitively, \verb|G_vi| is the named graph showing $G$'s state in version $i$ and \verb|G_ui| is another named graph showing the triples inserted into or deleted from $G$ by update $i$.
An example illustration is given
in Figure \ref{fig:example}. 

The provenance graph of named graph \verb|G| includes several kinds of nodes and
edges:
\begin{compactitem}
\item\verb|G_vi upd:version G_vi+1| edges that show the sequence
  of versions.  Whenever a \verb|upd:version| link is added between  \verb|G_vi| and  \verb|G_vi+1|, a backlink called \verb|upd:prevVersion| between  \verb|G_vi+1| and \verb|G_vi|;
\item nodes \verb|u1|,\ldots,\verb|un| representing the updates that
  have been applied to $G$, along with a \verb|upd:type| edge linking
  to one of \verb|upd:insert|, \verb|upd:delete|, \verb|upd:load|,
  \verb|upd:clear|, \verb|upd:create|, or \verb|upd:drop|.
 \item For all updates except create, an \verb|upd:input|
   edge linking \verb|ui| to \verb|G_vi|.
 \item For all updates except drop, an 
   \verb|upd:output| edge linking \verb|ui| to \verb|G_vi+1|.
 \item For insert and delete updates, an edge \verb|ui upd:data G_ui|
   where \verb|G_ui| is a named graph containing the triples that were
   inserted or deleted by \verb|ui|.
 \item Edges \verb|ui upd:source n| linking each update to each named
   graph \verb|n| that was consulted by \verb|ui|.  For an insert or
   delete, this includes all graphs that were consulted while
   evaluating $P$ (note that this may only be known at run time);
   for a load update, this is the name of the graph whose contents
   were loaded; create, drop and clear updates have no sources.
 \item Additional edges from \verb|ui| providing metadata for the
   update (such as author, commit time, log message, or the source
   text of the update); possibly using a standard vocabulary such as
   Dublin Core, or using OPM or PROV vocabulary terms.
\end{compactitem}

Note that this representation does not directly link triples in a
given version to places from which they were ``copied'' as it only contains the triples directly concerning the update in the history record.  However, each history record in combination with the rest of the history records in the provenance graph does provide enough information to recover previous versions on request.  As we store the source text of the update statements performed by each update in each history record of the provenance graph, we can trace backwards through the update sequence to  to identify places where triples were inserted or copied into or deleted from the graph. For queries, we consider a simple form of provenance which calculates a set of named graphs ``consulted'' by the query.  The set of sources of a pattern or query is computed as follows:
\begin{eqnarray*}
  \PSrc{C}{\D} &=& \bigcup\{\names_\mu(C) \mid  \mu \in \PEval{C}{\D}\}\\
 \PSrc{P_1 \andq P_2}{\D} &=& \PSrc{P_1}{\D} \cup
  \PSrc{P_2}{\D}\\
  \PSrc{P_1 \unionq P_2}{\D} &=& \PSrc{P_1}{\D} \cup
  \PSrc{P_2}{\D}\\
  \PSrc{P_1 \optq P_2}{\D} &=& \PSrc{P_1}{\D} \cup
  \PSrc{P_2}{\D}\\
  \PSrc{P \filterq R}{\D} &=& \PSrc{P}{\D} \\
\QSrc{\selectq{\var{\vec{X}}}{P}}{\D} &=& \PSrc{P}{\D}\\
\QSrc{\constructq{C}{P}}{\D} &=& \PSrc{P}{\D}
\end{eqnarray*}
where the auxiliary function $\names_\mu(C)$ collects all of the graph
names occurring in a ground basic graph pattern $C$:
\begin{eqnarray*}
  \names_\mu(\{t_1 ,\ldots,t_n\})  &=& \{\kw{DEFAULT}\}\\
\names_\mu( \graphq{A}{t_1,\ldots,t_n} ) &=& \{\mu(A)\}\\
\names_\mu(C  ~C') &=& \names_\mu(C) \cup \names_\mu(C')
\end{eqnarray*}
Here, we use the special identifier $\kw{DEFAULT}$ as the name of the
default graph; this can be replaced by its URI.

The $\QSrc{P}{\D}$ function produces a set $S$ of graph identifiers
such that replaying $\Eval{Q}{{\D|_S}} = \Eval{Q}{\D}$, where
$\D|_S$ is $\D$ with all graphs not in $S$ set to $\emptyset$
(including the default graph if $\kw{DEFAULT} \notin S$).
Intuitively, $S$ identifies graphs that ``witness'' $Q$, analogous to
why-provenance in databases~\cite{buneman01icdt}. This is not
necessarily the \emph{smallest} such set; it may be an
overapproximation, particularly in the presence of $P_1 \optq P_2$
queries~\cite{theoharis11ieee}.  Alternative, more precise notions of
source (for example involving triple-level annotations~\cite{flouris09iswc}) could also be used.

We define the provenance of an atomic update by translation to a
sequence of updates that, in addition to performing the requested
updates to a given named graph, also constructs some auxiliary named
graphs and triples (provenance record) in a special named graph  for provenance information called $prov$ (the provenance graph). We apply this translation to each update posed by the user, and execute the resulting updates directly without further translation. We detail how provenance information should be attached to each SPARQL Update operation. 

We consider simple forms of insert and delete operations that
target a single, statically known, named graph $g$; full SPARQL
Updates including simultaneous insert and delete operations can also
be handled.  In what follows, we write
``(metadata)'' as a placeholder where extra provenance metadata
(e.g. time, author, etc. as in Dublin Core or further information given by the PROV vocabulary \cite{prov-ontology}) may be added.
 DROP commands simply end the provenance collection, but previous versions of the graph should still be available.
\begin{itemize}
\item A graph creation of a new graph $\creategraph{g}$ 
is translated to 
\[
\begin{array}{l}
  \creategraph{g};\\
  \creategraph{g\_v_0};\\
  \insertdata{\graphq{prov}{\\
      \quad \rdf{g}{\mathtt{version}}{g\_v_0}, \rdf{g}{\mathtt{current}}{g\_v_0},\\
      \quad \rdf{u_1}{\mathtt{type}}{\mathtt{create}},\rdf{u_1}{\mathtt{output}}{g\_v_0},\\
      \quad \rdf{u_1}{\mathtt{meta}}{m_i}, (\text{metadata})\\
}}
\end{array}
\]
\item A drop operation (deleting a graph) $\dropgraph{g}$
is handled as follows, symmetrically to creation:
\[
\begin{array}{l}
  \dropgraph{g};\\
  \deletewhere{\graphq{prov}{\rdf{g}{\mathtt{current}}{g\_v_i}}};\\
  \insertdata{\graphq{prov}{\\
     \quad \rdf{u_i}{\mathtt{type}}{\mathtt{drop}},\rdf{u_i}{\mathtt{input}}{g\_v_i},\\
      \quad \rdf{u_i}{\mathtt{meta}}{m_i},(\text{metadata})\\
}}
\end{array}
\]
where $g\_v_i$ is the current version of $g$.
Note that since this operation deletes $g$, after this step the URI
$g$ no longer names a graph in the store; it is possible to create a
new graph named $g$, which will result in a new sequence of versions
being created for it.  The old chain of  versions will still be linked
to $g$ via the $\mathtt{version}$ edges, but there will be a gap in
the chain.
\item A clear graph operation $\clear{g}$
is handled as follows:
\[
\begin{array}{l}
  \clear{g};\\
  \deletewhere{\graphq{prov}{\rdf{g}{\mathtt{current}}{g\_v_i}}};\\
  \insertdata{\graphq{prov}{\\
      \quad \rdf{g}{\mathtt{version}}{g\_v_{i+1}}, \rdf{g}{\mathtt{current}}{g\_v_{i+1}},\\

      \quad \rdf{u_i}{\mathtt{type}}{\mathtt{clear}},\rdf{u_i}{\mathtt{input}}{g\_v_i},\\
      \quad \rdf{u_i}{\mathtt{output}}{g\_v_{i+1}},\rdf{u_i}{\mathtt{meta}}{m_i},\\
      \quad (\text{metadata})\\
}}
\end{array}
\]

\item A load graph operation $\loadinto{h}{g}$ 
is handled as follows:
\[
\begin{array}{l}
  \loadinto{h}{g};\\
  \deletewhere{\graphq{prov}{\rdf{g}{\mathtt{current}}{g\_v_i}}};\\
  \insertdata{\graphq{prov}{\\
      \quad \rdf{g}{\mathtt{version}}{g\_v_{i+1}}, \rdf{g}{\mathtt{current}}{g\_v_{i+1}},\\
      \quad \rdf{u_i}{\mathtt{type}}{\mathtt{load}}, \rdf{u_i}{\mathtt{input}}{g\_v_i},\\
      \quad \rdf{u_i}{\mathtt{output}}{g\_v_{i+1}}, \rdf{u_i}{\mathtt{source}}{h_j},\\
     \quad \rdf{u_i}{\mathtt{meta}}{m_i},(\text{metadata})\\
}}
\end{array}
\]
where $h_j$ is the current version of $h$. Note that a load will not create any new graphs because both the source and target should already exist. If no target exists, a new graph is created as outlined above with using the create operation.

\item An insertion $\insertinto{\graphq{g}{C}}{P}$ 
is translated to a sequence of updates that creates a new version and
links it to URIs representing the update, as well as links to the
source graphs identified by the query provenance semantics and a named graph containing
the inserted triples:
\[
\begin{array}{l}
  \creategraph{g\_u_i};\\
  \insertinto{\graphq{g\_u_i}{C}}{P};\\
  \insertinto{\graphq{g}{C}}{P};\\
  \creategraph{g\_v_{i+1}};\\
  \loadinto{g}{g\_v_{i+1}};\\
  \deletedata{\graphq{prov}{\rdf{g}{\mathtt{current}}{g\_v_i}}};\\
  \insertdata{\graphq{prov}{\\
\quad \rdf{g}{\mathtt{version}}{g\_v_{i+1}}, \rdf{g}{\mathtt{current}}{g\_v_{i+1}},\\
\quad \rdf{u_i}{\mathtt{input}}{g\_v_{i}}, \rdf{u_i}{\mathtt{output}}{g\_v_{i+1}},\\
\quad \rdf{u_i}{\mathtt{type}}{\mathtt{insert}}, \rdf{u_i}{\mathtt{data}}{g\_u_i} \\
\quad \rdf{u_i}{\mathtt{source}}{s_1},\ldots, \rdf{u_i}{\mathtt{source}}{s_m},\\
\quad \rdf{u_i}{\mathtt{meta}}{m_i}, (\text{metadata})
}}
  \end{array}
\]
where $s_1,\ldots,s_m$ are the source graph names of $P$.
\item A deletion $\deletefrom{\graphq{g}{C}}{P}$
is handled similarly to an insert, except for the update type
annotation.
\[
\begin{array}{l}
  \creategraph{g\_u_i};\\
  \insertinto{\graphq{g\_u_i}{C}}{P};\\
  \deletefrom{\graphq{g}{C}}{P};\\
  \creategraph{g\_v_{i+1}};\\
  \loadinto{g}{g\_v_{i+1}};\\
  \deletedata{\graphq{prov}{\rdf{g}{\mathtt{current}}{g\_v_i}}};\\
  \insertdata{\graphq{prov}{\\
\quad \rdf{g}{\mathtt{version}}{g\_v_{i+1}}, \rdf{g}{\mathtt{current}}{g\_v_{i+1}},\\
\quad \rdf{u_i}{\mathtt{input}}{g\_v_{i}}, \rdf{u_i}{\mathtt{output}}{g\_v_{i+1}},\\
\quad \rdf{u_i}{\mathtt{type}}{\mathtt{delete}}, \rdf{u_i}{\mathtt{data}}{g\_u_i}\\
\quad \rdf{u_i}{\mathtt{source}}{s_1},\ldots, \rdf{u_i}{\mathtt{source}}{s_m},\\
\quad \rdf{u_i}{\mathtt{meta}}{m_i}, (\text{metadata})
}}
  \end{array}
\]
Note that we still insert the deleted tuples into the $g\_u_i$.
\item The $\deleteinsert{C}{C'}{P}$ update can be handled as a
  delete followed by an insert, with the
  only difference being that both update steps are linked to the same metadata.
\item The $\copyto{h}{g}$, $\moveto{h}{g}$, and $\addto{h}{g}$ commands can be handled similarly to $\loadinto{h}{g}$; the only subtlety is that if $g = h$ then these operations have no visible effect, but the provenance record should still show that these operations were performed.
\end{itemize}

Our approach makes a design decision to treat
$\deleteinsert{C}{C'}{P}$ as a delete followed by an insert.  In
SPARQL Update, the effect of a combined delete--insert is not the same
as doing the delete and insert separately, because both phases of a delete--insert are evaluated against the same data store
before any changes are made.  However, it is not clear that this
distinction needs to be reflected in the provenance record; in
particular, it is not needed to ensure correct reconstruction.
Moreover, the connection between the delete and insert can be made
explicit by linking both to the same metadata, as suggested above.
Alternatively, the deletion and insertion can be treated as a single
compound update, but this would collapse the distinction between the
``sources'' of the inserted and deleted data, which seems undesirable for use-cases such as version control. 

Also note that our method does not formally take into account tracking the provenance of inferences. This is because of the complex interactions between SPARQL Update and the large number of possible (RDFS and the many varieties of OWL and OWL2) inference mechanisms and also because, unlike other research in the area \cite{flouris09iswc}, we do not consider it a requirement or even a desirable feature that inferences be preserved between updates. It is possible that an update will invalidate some inferences or that a new inference regime will be necessary.  A simple solution would be that if the inferences produced by a reasoning procedure are necessary to be tracked with a particular graph, the triples resulting from this reasoning procedure should be materialized into the graph via an insert operation, with the history record's metadata specifying instead of a SPARQL Update statement the particular inference regime used. We also do not include a detailed treatment of blank nodes that takes their semantics as existential variables, as empirical research has in general shown that blank nodes are generally used as generic stable identifiers rather than existential variables, and thus can be treated as simply minting unique identifiers \cite{Halpin:Interesting}.

\section{Update Provenance Vocabulary}

For the provenance graph  itself, we propose the following lightweight vocabulary called the ``Update Provenance Vocabulary'' (UPD) given in Table~\ref{tab:prov}.  Every time there is a change to a provenance-enabled graph by SPARQL Update, there is the addition of a provenance record to the provenance graph using the UPD vocabulary, including information such as an explicit time-stamp and the text of the SPARQL update itself.  Every step in the transaction will have the option of recording metadata using W3C PROV vocabulary (or even some other provenance vocabulary like OPM) explicitly given by the ``meta'' link in our vocabulary and semantics, with UPD restricted to providing a record of the `cut-and-paste' operations needed for applications of dynamic provenance like version control. We align the UPD vocabulary as a specialization of the W3C PROV vocabulary. A graph (\verb|upd:graph|) is a subtype of \verb|prov:Entity| and an update of a graph (\verb|upd:update|) is a subtype of \verb|prov:Activity|. For inverse properties, we use the inverse names recommended by PROV-O \cite{prov-ontology}. 
\begin{figure*}
 \centering 
\begin{tabular}{|c|p{7cm}|l|}
\hline
Name  &  Description  & PROV Subtype\\
\hline
upd:input  &     Link to provenance record from graph before an update
& prov:wasUsedBy\\
\hline
upd:output   &   Link from provenance record to a graph after update & prov:generated\\
\hline
upd:data & Changed data in insert/delete operation & prov:wasUsedBy\\
\hline
upd:version & Sequential link forward in time between a version of a graph and an update & prov:hadRevision \\
\hline
upd:prevVersion & sequential link backwards in time between a version of a graph and an update & prov:wasRevisionOf \\
\hline
upd:type  &  Type of update operation (insert, delete, load, clear, create, or drop) & prov:type\\
\hline
upd:current  & Link to most current state of graph & prov:hadRevision \\
\hline
upd:source  & Any other graph that was consulted by the update &  prov:wasUsedBy \\
\hline
upd:meta & Link to any metadata about the graph & rdfs:seeAlso\\
\hline
upd:user & User identifier (string or URI) & prov:wasAttributedTo \\ 
\hline
upd:text & Text of the SPARQL Update Query & prov:value \\
\hline
upd:time & Time of update to the graph & prov:atTime \\
\hline
\end{tabular}
\caption{Lightweight Update Provenance Vocabulary (UPD)}\label{tab:prov}
\end{figure*}

\section{Implementation Considerations}
 
So far we have formalized a logical model of the provenance of a graph as it evolves over time (which allow us to derive its intermediate versions), but we have not detailed how to store or query the intermediate versions of a graph efficiently.  For any given graph one should likely store the most up-to-date graph so that queries on the graph in its present state can be run without reconstructing the entire graph.  One could to simply store the graph \verb|G_vi| resulting from each update operation in addition to the provenance record, but this would lead to an explosive growth in storage requirements. This would also be the case even for the provenance graph if the storage of an auxiliary graph  \verb|G_ui| in a provenance record involved many triples, although we allow this in the UPD vocabulary as it may be useful for some applications. For those operating over large graphs, the contents of the named graphs  \verb|G_ui| that store inserted or deleted triples can be represented more efficiently by just storing the original graph and the SPARQL Update statements themselves in each provenance record given by the \verb|upd:text| property, and not storing the auxiliary named graphs given by \verb|upd:data|. 

Strategically, one can trade computational expense for storage in provenance, due to the immutability of the provenance information. A hybrid methodology to ameliorate the cost of reconstruction of the version of a graph  would be to store the graph at various temporal intervals (i.e. ``snapshots'').  For small graphs where storage cost is low and processing cost is high, it would make more sense to store all provenance information for the entire graph. In situations where the cost of processing is high and storage cost is low, storing the SPARQL Updates and re-running them makes sense to reconstruct the graph. In this case, it also makes sense to store ``snapshots'' of the graph at various intervals to reduce processing cost. Simulation  results for these scenarios are available.\footnote{\emph{http://www.ibiblio.org/hhalpin/homepage/notes/dbprov-implementation.pdf}}

\section{Conclusion}
Provenance is a challenging problem for RDF. By extending SPARQL Update, we have provided a method to use W3C PROV (and other metadata vocabularies) to keep track of the changes to triple-stores.  We formalized this approach by  drawing on similar work in database archiving and copy-paste provenance, which allow us to use SPARQL Update provenance records to reconstruct graphs at arbitrary instances in time. This work is a first step in addressing the important issue of RDF version control. We hope this will contribute to discussion of how to standardize descriptions of changes to RDF datasets, and even provide a way to translate changes to underlying (e.g. relational or XML) databases to RDF representations, as the same underlying ``cut-and-paste'' model has already been well-explored in these kinds of databases \cite{buneman06sigmod}. Explorations to adapt this work to the Google Research-funded DatabaseWiki project, and implementation performance with real-world data-sets is a next step \cite{buneman2011dbwiki}.  A number of areas for theoretical future work remain, including the subtle issue of combining it with RDFS inferences~\cite{flouris09iswc} or special-purpose SPARQL provenance queries~\cite{udrea10tocl,lopes10iswc}.

\paragraph*{Acknowledgements}
This work was supported in part by EU FP7 project DIACHRON (grant
number 601043).  The authors, their organizations and project funding partners are authorized to reproduce and distribute reprints and on-line copies for their purposes notwithstanding any copyright annotation hereon.

\bibliographystyle{plain}
\bibliography{iswc2014}

\end{document}